# Invited Discussion on "Randomization Tests to Address Disruptions in Clinical Trials: A Report from the NISS Ingram Olkin Forum Series on Unplanned Clinical Trial Disruptions"

Rachael V. Phillips[1,*] and Mark J. van der Laan[1]

[1] Division of Biostatistics, School of Public Health, University of California, Berkeley

* Corresponding author: Rachael V. Phillips, School of Public Health at UC Berkeley, 2121 Berkeley Way, Berkeley, California, 94704. Email: rachaelvphillips@berkeley.edu.

## 1   OVERVIEW

Disruptions in clinical trials may be due to external events, such as pandemics, warfare, and natural disasters. Resulting complications might include site closures, supply chain interruptions, and travel restrictions, and lead to unforeseen intercurrent events (events that occur after treatment initiation and affect the interpretation of the clinical question of interest and/or the existence of the measurements associated with it). In Uschner et al. (2023), the randomization test is presented as a strategy to address clinical trial disruptions. Inspired by Van Lancker et al. (2023), both Uschner et al. (2023) and Van Lancker et al. (2023) consider the guiding principles set forth in The International Council for Harmonisation of Technical Requirements for Pharmaceuticals for Human Use (ICH) E9(R1) and discuss how trial disruptions can be incorporated as intercurrent events in the statistical analysis plan (SAP) (ICH E9(R1), 2021). Uschner et al. (2023) draws on the hypothetical estimand strategies introduced in Van Lancker et al. (2023) for the causal intention-to-treat (ITT) estimand. Causal estimands are defined in terms of counterfactual outcomes, and to express them as a statistical estimand that can be estimated from data, some understanding of causal inference frameworks and identification results is necessary (Hernán and Robins, 2020; Pearl, 2009). Van Lancker et al. (2023) provides an overview of causal inference and missing data methodologies; therefore, we recommended reviewing this in advance of Uschner et al. (2023).

In Uschner et al. (2023), several example clinical trial disruptions are described, including treatment effect drift, population shift, change of care, change of data collection, and change of availability of study medication. A complex setting is presented — a randomized controlled trial (RCT) with (i) planned intercurrent events and (ii) unplanned intercurrent events and other complications brought on by external disturbances — and the clinical question of interest corresponds to the causal ITT estimand. Randomization tests are then presented as a means for non-parametric inference that is robust to violations of assumptions typically made in clinical trials. These assumptions are not explicitly mentioned, however. More generally, we do not see where the authors make the case that one should be going for a randomization test in a disrupted RCT with planned and unplanned intercurrent events. Even in the case where an external disruption does not occur, it is not clear how the randomization test is useful in an RCT with planned intercurrent events, while Targeted Learning estimation methods are valid in such settings (Gruber et al., 2022a,b; Gruber et al., 2023). The randomization test is limited in its applicability and "poor experimental design may render it not useful" (Rosenberger, 2019, p. 28; Louis, 2019). *We therefore request in the authors' rejoinder a clear theoretical demonstration in specific examples in this setting (RCT with planned and unplanned intercurrent events) or the simpler setting (RCT with planned intercurrent events) that a randomization test is the only valid inferential method relative to an estimation method following the Targeted Learning Roadmap*.

In this invited discussion, we comment on the appropriateness of Targeted Learning (TL) and the accompanying TL Roadmap in the context of disrupted clinical trials. We also highlight a few key articles related to the broad applicability of TL for RCTs and real-world data (RWD) analyses with planned and/or unplanned intercurrent events. We begin by introducing TL and motivating its utility in Section 2, and in Section 3, we provide a brief overview of the TL Roadmap for causal inference. In Section 4, we recite the example clinical trial disruptions presented in the Introduction of Uschner et al. (2023), and then discuss considerations and solutions based on the principles of TL. Concluding remarks are then given in Section 5.

## 2   THE CASE FOR TARGETED LEARNING TO ADDRESS DISRUPTIONS IN CLINICAL TRIALS

There are several parallels between learning from real-world data (RWD) and learning from disrupted clinical trial data. They both necessitate acknowledgement of potential sources of bias and specification of appropriate mitigation strategies. As discussed in Uschner et al. (2023) and Van Lancker et al. (2023), these data are subject to complexities (such as time-dependent confounding, censoring, competing risks, and positivity assumption violations), and their utility hinges on the quality of the analysis. To obtain sound scientific conclusions, it is therefore critical to consider robust analytic strategies in the SAP. Traditional, restrictive modeling assumptions cannot be realistically imposed in these settings. Machine learning algorithms represent more flexible, reasonable alternatives; however, they do not yield valid statistical inference (i.e., p-values and confidence intervals). Targeted Learning (TL), a form of causal machine learning and subfield of statistics, overcomes these challenges, leveraging machine learning for flexible estimation while implementing technical corrections required for valid statistical inference (van der Laan and Rose, 2011; van der Laan and Rose, 2018; Gruber et al., 2022a,b; Gruber et al., 2023a).

TL enables researchers to harness RWD as well as RCT data. The TL Roadmap provides a guide for statistical and causal inference with TL, and promotes the guiding principles and practices set forth in ICH E9(R1) addendum. TL is therefore readily applicable to handle RCT disruptions. FDA-funded initiatives have supported several TL projects (Gruber et al., 2022a,b,c; Gruber et al., 2023; Phillips et al. 2023). In Gruber et al. (2023) "Targeted learning: Towards a future informed by real-world evidence" and Gruber et al. (2022b) "Evaluating and improving real-world evidence with Targeted Learning", we discuss how the TL Roadmap provides step-by-step guide for generating valid evidence and assessing its reliability. Furthermore, and relevant to RCT disruptions, the roadmap also facilitates the development of alternative causal questions with better support in the data, which produce higher quality and more interpretable real-world evidence (RWE) (Gruber et al., 2022b). In Gruber et al. (2022a) "Developing a Targeted Learning-based Statistical Analysis Plan", we demonstrate how TL informs different aspects of SAP development, including explicit formulation of the role of intercurrent events in the definition of the estimand. We believe these three articles may be particularly helpful for to the authors to review for their rejoinder.

## 3   OVERVIEW OF THE TARGETED LEARNING ROADMAP FOR CAUSAL INFERENCE

Major guiding principles of TL are to (i) define the estimand *before* specifying the estimation procedure and (ii) use the general framework of targeted minimum loss-based estimation (TMLE) with super learning (SL) for estimation and inference of the estimand (van der Laan and Rubin, 2006; van der Laan et al., 2007; van der Laan and Gruber, 2012; Gruber et al., 2022a,b; Gruber et al., 2023). This estimation framework has been referred to as "the current frontier for embedding machine learning in the construction of doubly robust estimators" (Van Lancker et al., 2023, p. 104). It aligns with the ICH E9(R1) guidance, enabling "proper trial planning that clearly distinguishes between the target of estimation (trial objective, estimand), the method of estimation (estimator), the numerical result (estimate), and a sensitivity analysis" (ICH E9(R1), 2021, p. 4).

When the scientific question of interest is causal, the TL Roadmap incorporates the field of causal inference. Both a causal estimand and a statistical estimand need to be established; the latter identifies the former from the observed data, requiring identification assumptions to express the function defined in terms of counterfactual outcomes (causal estimand) as a function of the observed data distribution (statistical estimand). We typically use the general longitudinal G-computation formula to identify the post-intervention distribution under identification assumptions on the intervention nodes in the structural causal model (Robins, 1986; Snowden et al., 2011; Gruber and van der Laan, 2012). Additionally, the TL Roadmap incorporates a step for carrying out sensitivity analyses that evaluate the impact of identification assumption violations on inference for the causal estimand (Díaz and van der Laan, 2013; Gruber et al., 2022a,b; Gruber et al., 2023; Dang et al., 2023; Ho et al., 2023).

### 3.1  Defining the Estimand

Consider a clinical study aiming to investigate the causal ITT effect of a treatment assigned at baseline on an outcome of interest for a specified target population, like Van Lancker et al. (2023)

and Uschner et al. (2023). Additionally consider the motivating examples in Van Lancker et al. (2023) and Uschner et al. (2023), in which data were collected at baseline and then at eight future time points ($t = \{0, ... ,8\}$). Generally, we define the observed data structure in such settings as $O = (L_0, A_0, L_1, A_1, ..., L_8, A_8, Y)$, specifying the following:

- An outcome of interest measured at the final time point, $Y$.

- Intervention nodes $\{A_0, A_1, A_2, A_3, A_4, A_5, A_6, A_7, A_8\}$, where $A_0$ denotes treatment assignment at baseline, and $A_1$ through $A_8$ denote indicators of censoring/drop-out at and *intervention intercurrent events* (the subset of intercurrent events for which the intervention of preventing them from occurring is desired) at times $t > 0$.

- Covariates $L_t$, with $L_0$ denoting baseline covariates collected upon treatment randomization at baseline and $L_1$ through $L_8$ denoting time-varying covariates at times $t > 0$, including *non-intervention intercurrent events* (the sub-vector of intercurrent events for which no intervention is desired). We also note that the covariates $L_t$ collected at times $t < 8$ might include measurements of the outcome variable, i.e., these measurements are taken during regular follow-up times but before the final time point of interest.

We then define the population-level summary, the causal estimand, by contrasting post-intervention distributions of $Y$ under specified interventions on the intervention nodes. This framework, involving defining the intervention nodes among the longitudinal data elements and interventions on them, provides a general strategy for defining causal estimands and, through the G-computation formula and identification assumptions, corresponding statistical estimands. It naturally encompasses the strategies for addressing intercurrent events provided in the ICH E9(R1) guidance. For instance, intercurrent events addressed via hypothetical strategy or while-on-treatment strategy would be intervention intercurrent events, whereas those addressed via treatment policy strategy or composite variable strategy would be non-intervention intercurrent events.

To define the statistical estimand, we use the longitudinal G-computation formula for identification results, making the following identification assumptions on the intervention nodes:

- An extension to longitudinal settings of the missing at random assumption, the *sequential randomization assumption* states that the intervention nodes of interest at time $t$ are conditionally independent of the counterfactual outcome, given the observed history. A weaker SRA might be sufficient under known conditional independencies (See Section 2 of Gruber et al. 2022).

- An extension to longitudinal settings of the positivity assumption, the *sequential positivity assumption* states that within covariate strata, subjects have a nonzero probability of receiving each level of the interventions $A$ and $A(t)$. "The positivity assumption must hold at each time point with respect to the cumulative product of the conditional probabilities associated with following the intervention of interest at each time $t$. The probability of remaining uncensored can be 1 since no intervention of interest involves intervening to impose censoring." (Gruber et al., 2022, p. 4).

- "*Consistency*, an assumption that the outcome observed under the assigned treatment is equivalent to the counterfactual outcome under that level of exposure." (Gruber et al., 2022, p. 2).

Assuming the outcome assessment and record keeping is accurate, the consistency assumption is guaranteed for an ITT analysis. Randomization of the treatment $A_0$ guarantees the positivity and randomization assumptions are met, but only at baseline $t = 0$. The sequential randomization assumption is violated when there is unmeasured confounding of the associations between treatment, censoring, and the outcome (Gruber et al., 2023). For instance, if participant dropout is informed by an unmeasured variable that is also predictive of the future outcome, then we have a violation of the sequential randomization assumption. We also note that the choice of the intervention intercurrent events may affect the validity of these identification assumptions, so to ensure the causal estimand of interest can be identified from the observed data, identification assumptions should be considered when deciding on the subset of intercurrent events warranting an intervention in the structural causal model. For example, in a survival setting, if deaths related to the primary outcome are considered as intervention intercurrent events, this may lead to violations of the sequential randomization and positivity assumptions.

Careful consideration of the required identification assumptions facilitates an iterative process for defining the estimand. The validity of a causal interpretation of the study finding depends on how well identification assumptions are met. Additionally, violations / near violations of the positivity assumption threaten one's ability to reliably estimate the treatment effect from the data. As stated in the ICH E9(R1) guidance, "an iterative process should be used to reach an estimand that is of clinical relevance for decision-making, and for which a reliable estimate can be made" (ICH E9(R1), 2021, p. 11).

Considering the importance of identification assumptions and the iterative process for defining the estimand, we recommend such an elaboration by the authors. The example and causal estimands presented in Sections 3 and 4 of Uschner et al. (2023) provide the context to discuss whether the proposed causal estimands can be identified from the observed data example. Related to this, the causal ITT estimands defined in Section 4 of Uschner et al. (2023) should be accompanied by statistical ITT estimands that are functions of the observed data described in the motivating example of Uschner et al. (2023). Furthermore, the ICH E9(R1) guidance states that, "discontinuation of randomized treatment represents an intercurrent event to be addressed in the precise specification of the trial objective through the estimand" (ICH E9(R1), 2021, p. 2).; however, we did not see mention of the pre-pandemic/planned intercurrent events defined in the presentation of the causal estimands in Uschner et al. (2023). The second case study in Section 5 of Gruber et al. (2023) establishes a causal ITT estimand and subsequent statistical estimand that explicitly acknowledge the potential for intercurrent events at some time $t > 0$. This material may be helpful when considering identification of the causal ITT estimand from the observed data proposed in Uschner et al. (2023).

### 3.2 Defining the Estimation and Inferential Procedures

Once the statistical estimand is defined, we can develop a TMLE for it. This involves targeted estimation of the conditional densities of time-varying confounders across time and the outcome of interest, given the corresponding observed history, and an inference method. Please refer to Gruber et al. (2022a) for an approachable, more detailed overview of TMLE and additional references. The TMLE needs to be a priori specified in the SAP, but it can possibly be informed outcome-blind simulations, i.e., simulations that are not based on information regarding the observed treatment-outcome relationship (Gruber et al. 2022; Montoya et al. 2022; Ho et al. 2023; Dang et al. 2023).

### 3.3 Specifying Sensitivity Analyses

Lastly, sensitivity analyses should be specified a priori in the SAP. They will be used to obtain inference for the causal estimand under specified levels of causal gap due to violations of the sequential randomization assumption and potentially extra information on the plausible causal gap. Several examples of sensitivity analyses based on TMLE are provided in Diaz and van der Laan (2013), Gruber et al. (2022a,b), and Gruber et al. (2023).

## 4 CONSIDERATIONS FOR ADDRESSING DISRUPTIONS IN CLINICAL TRIALS WITH TARGETED LEARNING

Here, we draw on the examples of clinical trial disruptions presented in the Introduction of Uschner et al. (2023). We discuss considerations and solutions for each of them based on the principles of TL. In doing so, we investigate whether there appears to be a need to use the randomization test in any of these scenarios. We find out that we can handle these disruptions in a rather straightforward manner with TL.

### 4.1 Treatment Effect Drift

> Changes in patient's general health related to the pandemic shut down may lead to a drift in the patient populations baseline composition. This may result in initial trial assumptions being violated, as historical data is less relevant. For example, in a CNS trial studying a major depressive population, differences in the anticipated treatment effect may be smaller than initially anticipated with outside factors (shut-down, curfews, social isolation) take a toll on the mental state of the entire population. Similarly, outcome trials with mortality as an endpoint will have to address issues of confounding or competing outcomes, as more patients died due to COVID-19. This will reduce the overall study power. (Uschner et al., 2023, p. 1)

Suppose we collect a longitudinal data structure on each enrolled patient, involving baseline history and baseline treatment that is presumably randomized, and across monitoring times, we also collect time-dependent covariates and track outcome processes until the end of follow up. A priori one defines the manner in which patients are recruited and enroll in the study. The inclusion criterion, the process of enrollment, and the type of subjects enrolling will define the target population for which we aim to assess the causal impact of different treatment interventions.

The typical complexities longitudinal studies have is dropout due to intercurrent events that occur before observing the desired clinical outcome. The causal assumptions needed for identification of the counterfactual treatment-specific outcome distributions or survival curves are that the intervention nodes (such as treatment at baseline and indicator of drop-out at any point in time following baseline) are sequentially randomized, given the history at the time the intervention node is realized. If, due to a pandemic or another external event, the process of enrollment as well as the type of patients that enroll are changing, then that would not necessarily require a change of SAP, but it would change the interpretation of the estimand since it applies to the type of patients enrolled in the study. If the pandemic affects outcome distributions, then this would not necessarily change the estimand and proposed estimation procedure, but again, it might affect the interpretation of the estimand and power of the study.

The SAP needs to clarify how it deals with different types of intercurrent events, such as deaths. If they are viewed as drop-out/censoring events, then a pandemic might cause the study to have higher drop-out rates as anticipated. If the dropouts are caused by the pandemic, then they will satisfy the assumptions needed for identification. If they are viewed as a competing risk event, then the pandemic might cause a different outcome process than anticipated. Again, this might affect the power of the study.

Overall, we conclude that the described clinical trial disruptions might affect the interpretation of the results, the value of the true estimand, and the variance of the proposed estimator of the estimand, thereby affecting the power of the study. One way to resolve the power issue is to a priori specify that one will keep enrolling subjects until the variance estimator yields the desired power at the desired alternative. Due to the variance estimator being orthogonal to the t-statistic for the estimand, this type of adaptive sample size determination does not warrant penalization.

### 4.2 Population Shift

> In studies where the safety of a drug is not completely known, women of child-bearing age were discontinued from treatment during the pandemic, due to the inability of the investigator to ensure reliable pregnancy tests were being conducted prior to allowing the patient to continue in the trial. (Uschner et al., 2023, p. 1)

Typically, one defines discontinuation of treatment as a censoring event, so in this scenario, women of bearing age would be censored at that time. Given that the probability of dropping out at time $t$ is equal to one for women who are of child-bearing age at time $t$, the positivity assumption is theoretically violated. This assumption is required for identification of the desired causal effect. If one can argue that being a woman of child-bearing age is not a predictor of the clinical outcome, or that it is blocked by another confounder measured at the same time point, then one can still obtain identification with the G-computation formula that ignores this variable. However, the latter may be unlikely to hold, and an alternative strategy is therefore required. One such strategy is to change the target population to exclude women of child-bearing age at baseline.

In Petersen et al. (2012) "Diagnosing and responding to violations in the positivity assumption", this strategy is listed as one of many practical solutions: "when the subset of covariates responsible for positivity violations is low or one dimensional, such an approach [of restricting the sample] can be implemented simply by discarding subjects with covariate values not represented in all treatment groups." (Petersen et al., 2012, p. 20). Making this change to the enrollment criteria does not affect the validity of the statistical methods for inference with respect to the newly defined estimand. We recommend that this change be communicated to the regulatory agency before analyzing the outcome data. This change would still result in a loss of sample size. As mentioned above, the power issue could be handled with an adaptive sample size design that is based on tracking the variance estimators of the proposed estimator of the estimand across time.

### 4.3 Change of Care and Change of Data Collection

> Due to the war in Ukraine, patients in ongoing clinical trials were either withdrawn from the trials or transferred to other sites in neighboring countries, in cases where the patient was able to move there. In cases where a patient moved from one site to another, the internal reliability of the endpoint assessment may be questionable. For example, in situations where the treatment under study is administered on top of Standard of Care (SOC), what is considered SOC in one country is often very different from the SOC in another country. A patient fleeing from one country to another may be required to change the SOC. (Uschner et al., 2023, p. 1)

> During the COVID-19 shut down and the war in Ukraine, in-person patient visits were converted to phone visits (when possible). This leads to a lower/different data quality of the data collected from these patients and visits, versus patients that had access to in-person visits. For example, in a recent schizophrenia trial, assessments were to be made by well-trained site staff. These assessments included tasks in which the patient had to draw shapes, repeat patterns from memory and other visual and behavioral tasks. During the shutdown, when sites were closed to in-person visits, these assessments were done via video call. However, video calls are imperfect in terms of Wi-Fi quality, situational distraction, and general attention, making comparability of assessments done via video versus those done in person questionable. (Uschner et al., 2023, p. 2)

These changes to the study could affect the validity of the identification assumptions due to the measured time-dependent confounders not being sufficient for validity of the sequential randomization assumption. This would make the G-computation formula deviate from the desired post-intervention distribution. Failure to measure the desired outcome could be dealt with as missingness, but like the previous example, one must worry about deterministic patterns for particular subjects as this would violate the positivity assumption needed for identification of causal quantity and affects the power of the study.

If one can anticipate the potential need for a surrogate outcome as an alternative for the desired clinical outcome, then the recommendation would be to measure the surrogate outcome on all subjects, while the study aims to also measure the desired outcome. This would allow estimation of the causal effect on the surrogate outcome, but also allows one to still estimate the causal effect on the desired outcome. The subjects without the desired outcome measurement would still play an important role in the precision of the estimator of the causal effect on it, with precision gains increasing as the surrogate outcome's predictive power for the desired outcome increases (Rose and van der Laan, 2011).

Two-stage sampling designs are a natural way to deal with a situation where not all the desired data is collected, possibly due to unforeseen circumstances like a pandemic. In such designs, one randomly selects a subset of the patients among those that have a missing outcome, mismeasured outcome, or lack measurement of other key variables, and then they obtain the desired data from them using extra resources. Due to the random sampling being controlled and thereby understood, the resulting study still allows identification of the desired causal effect, while the subjects without all the desired data are still used to gain efficiency. If the random sampling of the subjects for which the desired outcome is measured is due to the pandemic, it might still be a well understood missingness process, not that different from one that is controlled by the experimenter. A TMLE for two-stage designs, which incorporates inverse probability of censoring weights (IPCW), was introduced in Rose and van der Laan (2011).

### 4.4 Change of Availability of Study Medication

> Investigative Medicinal Product (IMP) supply issues can lead to a disruption in the planned randomization of patients. The Interactive Response Technology (IRT)/ Interactive Voice Response System (IVRS) provides the sponsor with an option to include "forced randomization" within an IRT system. When this option is turned on, and a kit for that treatment arm that a patient is supposed to be randomized to (according to the next entry on the randomization list) is not available, the IRT system skips that entry and assigns the patient to a treatment kit that is actually available at the site. One option of "forced randomization" implemented in the IRT allows the skipped randomization assignments to then "back-filled" by patients at other sites that have that treatment kit available. (Uschner et al., 2023, p. 2)

This could be handled with an ITT intervention, where one evaluates the causal impact of the randomized treatment, even though for some subjects the randomized treatment is replaced by another treatment due to non-compliance or, in this case, lack of availability. Given that the non-compliance to the randomized treatment is due to an external event, it should not be informed by individual characteristics of the subject. As a result, this should not affect identification of the per-protocol treatment effect, but as always, unbiased estimation of such an estimand is more challenging than an ITT estimand.

## 5   SUMMARY

Clinical trial disruptions, such as a pandemic, could impact a trial in a number of ways: subject recruitment, impacting the sampling of units and thereby distribution of the observed data; the definition of treatment or censoring changes, e.g., perhaps there is extra censoring due to pandemic; the definition of the outcome; how key time-dependent confounders are measured and how regularly we measure them. As we saw in Section 4, in many cases, we just accept the statistical estimand with the altered data and definitions, and thus do not change the TMLE and inference method. We do, however, acknowledge that the changes caused by the disruption will affect the interpretation of the estimand and the power of study. In some cases, we must define a new estimand or even change the design retroactively (e.g., by collecting extra data on random subset of subjects with a two-stage design). The latter might be necessary when the changes due to the unforeseen events affects identification of the causal estimand, further highlighting the utility of identification assumptions in defining a reliable estimand.

Overall, it is best to anticipate these intercurrent events in the a priori specified SAP and estimand (Gruber et al., 2023a). Some post-hoc changes in the SAP are allowable if they are not affected by outcome data and are only random through orthogonal data. By focusing on estimands that (i) incorporate intervening intercurrent events and (ii) utilize estimation procedures that can handle them, a robust SAP can be established that handles most planned and unplanned intercurrent events, possibly up to non-informative post-hoc modifications. We showed that with TL, a well-defined SAP based on the TL Roadmap with some post-hoc or a priori specified adaptations handles these challenges quite well; in particular, without a need for going for a randomization test.

We encourage further investigation on the utility of randomization tests in this complex setting: RCTs with (i) planned intercurrent events and (ii) unplanned intercurrent events and other complications brought on by external disturbances. We request from the authors in their rejoinder a clear theoretical demonstration in specific examples in this setting or the simpler setting (RCT with planned intercurrent events) that a randomization test is the only valid inferential method relative to an estimation method following the TL Roadmap.